\documentclass{aa}
\usepackage{graphicx}

\begin{document}

\title{A photometric mode identification method, including an improved
 non-adiabatic treatment of the atmosphere\thanks{The non-adiabatic 
 eigenfunctions needed 
 for the mode identification are available upon request from the authors.}}

\author{M.-A. Dupret \inst{1} \and J. De Ridder \inst{2} \and P. De Cat 
\inst{2}\thanks{Postdoctoral Fellow of the Fund for Scientific
Research, Flanders, Belgium}
 \and C. Aerts \inst{2} \and R. Scuflaire \inst{1} \and
A. Noels \inst{1} \and A. Thoul \inst{1}\thanks{Chercheur Qualifi\'e au Fonds
National de la Recherche Scientifique (Belgium).}}

\offprints{M.-A. Dupret, \\ \email{madupret@ulg.ac.be}}

\institute{Institut d'Astrophysique et de G\'eophysique de 
l'Universit\'e de Li\`ege, all\'ee du 6 Ao\^ut 17,
B-4000 Li\`ege, Belgium \and 
Instituut voor Sterrenkunde, Katholieke Universiteit
Leuven , Celestijnenlaan 200 B, 3001 Leuven, Belgium}

\date{Received / Accepted }

\titlerunning{An improved method of photometric mode identification}
\authorrunning{M.-A Dupret et al.}

\abstract{
%
We present an improved version of the method of photometric mode
identification of Heynderickx et al.~(\cite{hey}). 
Our new version is based on the inclusion of precise
non-adiabatic eigenfunctions determined in the outer stellar atmosphere
according to the formalism recently proposed by Dupret et al.\ (\cite{dup}).
Our improved photometric mode identification technique is 
therefore no longer dependent on {\it ad hoc\/}
parameters for the non-adiabatic effects.  It contains the complete physical
conditions of the outer atmosphere of the star, provided that rotation does
not play a key role. We apply our improved method to the two slowly 
pulsating B stars HD~74560 and HD~138764 and to the $\beta$ Cephei star 
EN (16) Lac. Besides identifying the degree $\ell$ of the pulsating
stars,  our method is also a tool for improving the knowledge of 
stellar interiors and atmospheres, by imposing constraints on parameters
such as the metallicity and the mixing-length parameter $\alpha$
(a procedure we label non-adiabatic asteroseismology).
}

\maketitle

\section{Introduction}  

A crucial problem in asteroseismology is mode identification.  Firstly
because, from a theoretical point of view, despite the linear non-adiabatic
predictions, the mode selection mechanisms are not well understood for many
kinds of pulsating stars ($\delta$ Scuti, $\beta$ Cephei, slowly pulsating B
stars, $\gamma$ Doradus, roAp stars,~\ldots). Secondly because, from an
observational point of view, we do not resolve the disks of stars 
other than the Sun so that we can only observe disk-integrated 
quantities. Thirdly, the rotational splittings
and the ``avoided crossing'' effect produce such a complicated 
power spectrum that a mode identification based on the frequencies 
alone is generally impossible.

Currently, both spectroscopic and photometric mode identification techniques
are being used. The latter methods are based on multi-colour photometry, and
are the subject of this paper.
The principle of these methods is to observe the photometric variations due to
stellar oscillations in different colours and compare them to the theoretical
predictions at the appropriate wavelengths (Dziembowski \cite{dziem2}, Stamford
\& Watson \cite{stam}, Watson \cite{wat}, Garrido et al. \cite{gar2},
Heynderickx et al.  \cite{hey}, Garrido \cite{gar}). However, all these 
methods have an important drawback: their theoretical predictions are 
very sensitive to the non-adiabatic temperature variations at the 
photosphere (Cugier et al. \cite{cug}, Balona \& Evers \cite{balev}).  

Dupret et al.\ (\cite{dup}) developed a non-adiabatic code including a detailed
treatment of the pulsation in the outer atmosphere.  In this paper, we show how
this non-adiabatic treatment opens the way to a significant improvement of the
discriminant power of mode identification methods based on multi-colour
photometry.  Indeed, by using Dupret et al.'s calculations, we are able to
eliminate the weakest point of the mode identification method, namely the 
{\it ad hoc\/} parameters to express the non-adiabaticity.  We present the results
obtained for two slowly pulsating B stars (SPBs) observed by De Cat \& Aerts
(\cite{decat}) and for the $\beta$ Cephei star EN (16) Lac.

\section{Monochromatic magnitude variations of a non-radially 
pulsating star}
 
Theoretical expressions for the monochromatic magnitude variations of a
non-radially pulsating star have been derived by different authors.  Dziembowski
(\cite{dziem2}) was the first to derive an expression for the bolometric
magnitude variation of a non-radially pulsating star. He suggested also
that a Wesselink technique could be formulated from these expressions. 
Balona \& Stobie (\cite{balsto}) recast the suggestion of Dziembowski in an 
observationally feasible way.
Stamford \& Watson (\cite{stam}) derived an expression for the monochromatic
magnitude variations of a non-radially pulsating star. They proposed to compute
the local emergent monochromatic flux variation on the base of equilibrium
atmosphere models (see Eq.\ (\ref{Fpert}) below) and they simplified the way to
compute the influence of the stellar surface distortion.  Watson (\cite{wat}) 
improved the method by taking the variation of the limb darkening into account 
(see Eq.\ (\ref{hpert}) below), and discussed the importance of the different 
terms in the equation giving the
monochromatic magnitude variations of a non-radially pulsating star.  Garrido et
al.\ (\cite{gar2}) and Garrido (\cite{gar}) derived a method of mode identification 
using Str\"omgren photometry, based on the formalism of Watson (\cite{wat}), and 
applied it to $\delta$ Scuti and $\gamma$ Doradus stars. Heynderickx et al. 
(\cite{hey}) derived an expression for the surface
distortion of a non-radially pulsating star in a Lagrangian formalism. He
developed a method of mode identification based on photometric amplitude ratios
and applied it to $\beta$ Cephei stars.

In all the previously cited papers, the non-adiabatic character of the pulsation
was neglected or treated with an {\it ad hoc\/} parameter.  Cugier et al.\
(\cite{cug}) were the first to use non-adiabatic computations for
photometric mode identification, using the non-adiabatic pulsation
code of Dziembowski (\cite{dziem}). The same code was also used by 
Balona \& Evers (\cite{balev}) for mode identifications of $\delta$ Scuti stars.
Townsend (\cite{tow}) used a non-adiabatic code for the photometric modelling
of SPBs.

It is useful to detail the assumptions made, sometimes implicitly, by the
previous authors and in our method, in order to derive an expression for the
monochromatic magnitude variations of a non-radially pulsating star. In sections 
\ref{ourmethod} and \ref{comparison}, we will put forward the improvement
and specificities of our method, compared to the one of the previous authors.
\begin{description}
\item[a1)] We work in the linear approximation.
\item[a2)] We neglect the coupling of modes due to the interaction 
  between rotation and pulsation.
  The angular dependence of a non-radial mode is thus described by
  a single spherical harmonic.
\item[a3)] We assume that the gas column of the atmosphere at a given 
angular position ($\theta$, $\phi$) is well described by a plane parallel 
atmosphere, which we call the {\it local atmosphere}. 
\item[a4)] For the geometrical distortion of the stellar surface, 
we work in the one-layer approximation. It is assumed that
the visible part of the star, i.e.~the photosphere, can be described by a single 
surface which is spherical at equilibrium. 
The radius $R_0$ of this sphere is the radius of the star, 
and in our method, we assume that it corresponds to the layer where the local 
temperature is equal to the effective temperature of the star.
During the pulsation, it is assumed that this surface follows the 
movement of the matter. The surface distortion can therefore be deduced
from the displacement field: $\vec{\xi}(R_0, \theta, \phi, t)$.
\item[a5)] We assume that, during the pulsation cycle, 
the monochromatic outwards flux $\vec{F_{\lambda}}^+$ of the local 
atmosphere is, for each given time, the same as the monochromatic outwards
flux of an equilibrium plane parallel atmosphere model.
\item[a6)] We assume that, during the pulsation cycle,
$\vec{F_{\lambda}}^+$ remains perpendicular to the photosphere.
\item[a7)] We assume that $\vec{F_{\lambda}}^+$ 
does not depend on the optical depth in the local atmosphere.
\item[a8)] We assume that the local atmosphere depends only on 
  two varying parameters: the local effective temperature and the 
  local effective gravity. The chemical composition of the local atmosphere
  is assumed to remain constant.
\item[a9)] We assume that during the pulsation cycle, 
the limb darkening law $h_{\lambda}$ of the local 
atmosphere is, for each given time, the same as the 
limb darkening law of an equilibrium plane parallel atmosphere model
with the orientation given by assumption (a4) and (a6). 
\end{description}

Assumption (a4) needs some precisions. We have shown in Dupret et al. (\cite{dup}) 
that the relative Lagrangian variation of the optical depth ($\delta\tau/\tau$) is not 
negligible in stellar atmospheres. However, on the basis of non-adiabatic computations,
we have checked that the relative difference 
between the displacement of constant optical depth layers and the ``real''  
displacement of the matter ($\delta\tau/(\kappa\,\rho\, R_0)$) is very small. 
For example, $|\delta\tau|/(\kappa\,\rho\,R_0)\simeq 0.005$ at the photosphere
of a typical $\beta$~Cephei model and for the fundamental radial mode, with
a relative radial displacement normalized to 1 at the photosphere.
Therefore assumption (a4) is appropriate for the determination of the geometrical
distorsion of the stellar surface (at least for g-modes and moderate order p-modes).
However, we showed in Dupret et al. (\cite{dup}) that it is not appropriate to 
assume that the Lagrangian variation of the temperature is equal to the variation
of the temperature at constant optical depth, because
$\partial\ln T/\partial\tau\,|\delta\tau|$ is not negligible compared to 
$|\delta T/T|$ at the photoshere.

We note that assumption (a7) concerns only the flux. We do not make this hypothesis
for the temperature which depends strongly on the optical depth in stellar atmospheres
(see Dupret et al. \cite{dup}, Fig. 1).

Assumptions (a5) and (a9) have the same physical justification as explained in
Dupret et al. (\cite{dup}). Because of the very small thermal relaxation time of
the atmosphere, we assume that, at each time of the pulsation cycle, the local
atmosphere remains in radiative equilibrium. Using this assumption, 
Dupret et al. (\cite{dup}) assumed that, during the pulsation cycle, the
temperature distribution ($T(\tau)$ law) in the local atmosphere was, for each
given time, the same as the temperature distribution of an equilibrium
atmosphere model.  For the same physical reasons, we make now the same
assumption for the local monochromatic outwards flux $\vec{F_{\lambda}}^+$ and
for the limb darkening law $h_{\lambda}$.

From assumptions (a5), (a7) and (a8), the monochromatic outward flux 
in the local atmosphere is given by:
{\setlength\arraycolsep{2pt}
\begin{eqnarray}  \label{Fpert0}
\left(F_{\lambda}^+\right)_0 &+& \delta F_{\lambda}^+(\theta, \phi, t) \nonumber \\ 
&=& F_\lambda^+\left[\,\left(T_{\rm eff}\right)_0 
+ \delta T_{\rm eff}(\theta, \phi, t) ,\,
g_0 + \delta g_{\rm e}(\theta, \phi, t)\,\right],
\end{eqnarray}} 
where $F_{\lambda}^+ = \left|\vec{F_{\lambda}}^+\right|$.
In the linear approximation, we have thus by perturbing Eq.\ (\ref{Fpert0}):
\begin{eqnarray} \label{Fpert}
\frac{\delta F_{\lambda}^+}{F_{\lambda}^+}
&=& \left(\frac{\partial \ln F_{\lambda}^+}{\partial \ln T_{\rm eff}}\right)\,
\frac{\delta T_{\rm eff}}{T_{\rm eff}}
\:+\:\left(\frac{\partial \ln F_{\lambda}^+}{\partial \ln g}\right)\,
\frac{\delta g_{\rm e}}{g_{\rm e}}
\nonumber \\
&\equiv&  \alpha_{T\lambda}\:\frac{\delta T_{\rm eff}}{T_{\rm eff}}
\:+\:\alpha_{g\lambda}\: \frac{\delta g_{\rm e}}{g_{\rm e}}\:.
\end{eqnarray}
Eq.\ (\ref{Fpert0}) was first proposed by Stamford \& Watson 
(\cite{stam}).
We proceed similarly for the variation of the monochromatic limb darkening.
From assumptions (a4), (a6), (a8) and (a9), we obtain in the linear 
approximation:
\begin{eqnarray} \label{hpert}
\frac{\delta h_{\lambda}}{h_{\lambda}}
&=& \left(\frac{\partial \ln h_{\lambda}}{\partial \ln T_{\rm eff}}\right)\,
\frac{\delta T_{\rm eff}}{T_{\rm eff}}
\:+\:\left(\frac{\partial \ln h_{\lambda}}{\partial \ln g}\right)\,
\frac{\delta g_{\rm e}}{g_{\rm e}}\\
&+&\left(\frac{\partial \ln h_{\lambda}}{\partial \mu}\right)\,
\delta\left(\mu\right)\:,
\end{eqnarray}
where $h_{\lambda}$ is the normalized limb darkening law
($\int_0^1 h_\lambda(\mu)\:\mu\:{\rm d}\,\mu\;=\;1$),
 $\mu=\vec{n}\cdot\vec{e_{z'}}$, $\vec{n}$ is the normal to the 
photosphere and $\vec{e_{z'}}$ is the unit vector pointing towards 
the observer. A similar equation was first proposed by Watson (\cite{wat}). 

We denote by $\epsilon$ the amplitude of relative radial displacement
in the photosphere:
\begin{equation} \label{xileg}
\xi_r(\theta, \phi, t) \;=\; R_0\,\epsilon\,P_\ell^m (\cos \theta)\,
\cos(\sigma\,t\,+\,m\,\phi)\:,
\end{equation}
where $\theta$ and $\phi$ are the usual spherical coordinates 
with respect to the polar (rotation) axis of the star, $P_\ell^m (x)$ 
is the associated Legendre function of degree $\ell$ and 
azimuthal number $m$ and $\sigma$ is the angular oscillation frequency.

The quantities $\delta T_{\rm eff} / T_{\rm eff}$ and $\delta g_{\rm e} / g_{\rm
e}$ can be computed by the non-adiabatic code of Dupret et al.\
(\cite{dup}).  Because the Eulerian variation of the gravitational potential at
the surface is always very small (~$|\psi'|\,\ll\,|g\,\xi_r|$~), it appears that
in very good approximation, $\delta g_{\rm e} / g_{\rm e}$ is in opposite phase 
with the radial displacement. These two quantities can be expressed in term of the
associated Legendre functions:
\begin{eqnarray} \label{TeB}
\frac{\delta T_{\rm eff}}{T_{\rm eff}} (\theta, \phi, t) &=&
f_T\:\epsilon\:P_\ell^m (\cos \theta)
\:\cos(\sigma\,t\,+\,m\,\phi\,+\,\psi_T)
\,, \\ \label{geD}
\frac{\delta g_{\rm e}}{g_{\rm e}} (\theta, \phi, t) &=&
-\,f_g\:\epsilon\:P_\ell^m (\cos \theta)\:\cos(\sigma\,t\,+\,m\,\phi)\:,
\end{eqnarray} 
where $f_T$ and $f_g$ are the amplitudes of $\delta T_{\rm eff} / T_{\rm eff}$ 
and $\delta g_{\rm e} / g_{\rm e}$ corresponding to a normalized radial
displacement at the photosphere.
On the basis of the previous assumptions and equations, an expression can
be derived for the monochromatic magnitude variation of a non-radially
pulsating star:

\begin{eqnarray} \label{deltaml}
\delta m_\lambda &=&
-\:\frac{2.5}{\ln 10}\;\epsilon\,\:P_\ell^m(\cos i)\:b_{\ell\lambda} 
\nonumber \\ 
& & \Big(\,-\:(\ell-1)(\ell+2)\:\cos(\sigma\,t) \nonumber \\ 
& & \hspace*{0.5cm}+\;f_T\:\cos(\sigma\,t + \psi_T)
\,\left(\alpha_{T\lambda}\,+\,\beta_{T\lambda}\,\right)
\nonumber \\
& & \hspace*{0.5cm}-\;f_g\:\cos(\sigma\,t)\,\left(\alpha_{g\lambda}\,+\,
\beta_{g\lambda} \,\right)\,\Big)\:,
\end{eqnarray}
where $\delta m_\lambda$ is the variation of the monochromatic 
magnitude as seen by the observer, $i$ is the inclination angle between the
stellar axis and the direction towards the observer and
\begin{eqnarray} 
b_{\ell\lambda} & \equiv & \int_0^1\,h_\lambda\:\mu
\:P_\ell\:{\rm d}\,\mu\:, \\
\beta_{T\lambda} & \equiv & 
  \frac{\partial\ln\,b_{\ell\lambda}}{\partial\ln T_{\rm eff}}\:, \\
\beta_{g\lambda} & \equiv & 
\frac{\partial\ln\,b_{\ell\lambda}}{\partial\ln g}\:.
\end{eqnarray}
For the derivations leading to Eq. (\ref{deltaml}), we refer to
Dziembowski (\cite{dziem2}), Stamford \& Watson (\cite{stam}),
Watson (\cite{wat}) and Heynderickx et al.\ (\cite{hey}).

In Eq.\ (\ref{deltaml}), the term proportional to $(\ell-1)(\ell+2)$ corresponds
to the influence of the stellar surface distortion, the term proportional to
$f_T$ corresponds to the influence of the local effective temperature variation
and the term proportional to $f_g$ corresponds to the influence of the effective
gravity variation.  In our applications, we computed the coefficients
$\alpha_{T\lambda}$ and $\alpha_{g\lambda}$ (derivatives of the monochromatic
flux) from the models of Kurucz (\cite{kur}).  An analytical law for the limb
darkening is needed for the computation of $b_{\ell\lambda}$ and its
derivatives.  For the present paper, we used a quadratic law (Wade \& Rucinski
\cite{wade}).  We note that an improved non-linear limb darkening law has been
proposed by Claret (\cite{clar}), but his computations were only done for
Str\"omgren filters, while our applications concern Geneva and Johnson
filters.

In multi-colour photometry, one observes the integral
of the monochromatic magnitude variation over the response of the filter:
\begin{equation} \label{filter}
\delta m_i\:=\:\frac{\int_{\lambda_{\rm min}}^{\lambda_{\rm max}}
\delta m_\lambda\,w_i(\lambda)\,{\rm d}\lambda}
{\int_{\lambda_{\rm min}}^{\lambda_{\rm max}}w_i(\lambda)\,{\rm d}\lambda}\:,
\end{equation}
where $w_i(\lambda)$ is the response curve of the filter $i$.
Therefore, the different terms of Eq. (\ref{deltaml}) depending on $\lambda$ 
have to be integrated, following Eq. (\ref{filter}).

\section{Our version of the mode identification method} \label{ourmethod}
 
The linear theory does not permit to predict the amplitudes of the
eigenfunctions. Therefore, it is appropriate to use amplitude ratios and phase
differences between different filters when comparing the theoretical predictions
to the observations. On one hand, for $\delta$ Scuti stars, the observations and
the theoretical predictions give significant phase-lags between the different
filters. Mode identification methods using these phase-lags have been proposed
by Garrido et al. (\cite{gar2}) and Balona \& Evers (\cite{balev}). On the other
hand, for $\beta$ Cephei stars, SPBs and $\gamma$ Doradus stars, no phase-lags
are observed (in agreement with the small phase-lags predicted by the theory).
For the latter stars, mode identification methods based on amplitude ratios are
thus appropriate (Heynderickx et al.\ \cite{hey}).  We adopt here such a
method. The theoretical procedure of our mode identification method is the
following:
\begin{enumerate}
\item We compute a stellar model with the appropriate effective temperature, 
luminosity and mass. In our applications, we used the new Code Li\'egeois
d'\'Evolution Stellaire written by one of us (R.\ Scuflaire).
\item We perform non-adiabatic computations for different degrees $\ell$ and for
pulsation frequencies close to the observed ones. In our applications, we used
the non-adiabatic code by Dupret et al.\ (\cite{dup}). These computations give
the coefficients $f_T$, $\psi_T$ and $f_g$ for different degrees $\ell$.
\item For each filter $j$ and for each $\ell$, we compute 
\begin{eqnarray} \label{ampl}
A_{j,{\rm th}} &=& |b_{\ell\,j}|\,\Big|\:(1-\ell)(\ell+2) \nonumber \\
& & \hspace*{0.7cm}+\; f_T\;e^{i \psi_T}
\left(\alpha_{T\,j}\,+\,\beta_{T\,j}\,\right)
\nonumber \\
& & \hspace*{0.7cm}-\;f_g\,\left(\alpha_{g\,j}\,+\,\beta_{g\,j}
\,\right)\;\;\;\Big|\:,
\end{eqnarray}
using the atmosphere models of Kurucz (\cite{kur}) to compute $\alpha_T$ and 
$\alpha_g$.
\item We choose a reference filter (indicated with subindex $1$).
For B stars, this reference filter is the U filter giving the highest amplitudes
and thus the highest S/N ratio. We compare the theoretical amplitude 
ratios $(A_{j,{\rm th}}/A_{1,{\rm th}})$ to
the observed amplitude ratios $(A_{j,{\rm obs}}/A_{1,{\rm obs}})$.  The
identified degree $\ell$ is the value which minimizes the $\chi^2$:
\begin{equation}
\sum_{j=2}^k\,\left[\,\frac{A_{j,{\rm th}}}{A_{1,{\rm th}}}\,-\,
                      \frac{A_{j,{\rm obs}}}{A_{1,{\rm obs}}}\,\right]^2,
\end{equation}    
where $k$ is the number of filters.

\end{enumerate}

We note that the non-adiabatic predictions depend on some dominant parameters of
the theoretical models (e.g.\ the metallicity for $\beta$ Cephei stars and SPBs,
the mixing length parameter $\alpha$ for $\delta$ Scuti and $\gamma\,$Doradus
stars).  Therefore, these parameters can be constrained by a feed-back process
after a unique mode identification is achieved. 
We call this feed-back process {\it non-adiabatic asteroseismology\/}, 
in which we iterate the procedure described above by
adjusting the stellar parameters for the identified mode, until we find the best
fit between theory and observations. We will illustrate this feed-back process
below for the estimation of the metallicity of the $\beta$ Cephei star EN Lac.

\section{Comparison with other methods} \label{comparison}

The difference between our method and the one proposed by other authors
is in the way of estimating the influence of the 
effective temperature variation and the effective gravity variation
in Eqs.~(\ref{deltaml}) and (\ref{ampl}).

\subsection{Mechanical boundary condition}

As a preliminary for the estimation of the relative effective temperature 
variation $\delta T_{\rm eff}/T_{\rm eff}$ and the relative effective gravity 
variation  $\delta g_e / g_e$, the first step of the procedure followed by 
Watson (\cite{wat}), Heynderickx et al. (\cite{hey}), 
Cugier et al. (\cite{cug}) and Balona \& 
Evers (\cite{balev}) was to compute $\delta P/P$ at the photosphere. As 
initially proposed by Buta \& Smith (1979), they used the following formula:
\begin{equation} \label{dPbut}
\frac{\delta P}{P} \;=\; \Big(\,\ell(\ell+1)\,K\:-\:4\:-\:K^{-1}\,\Big)
\,\frac{\xi_r}{R_0}\:,
\end{equation}
where $K$ (sometimes denoted by $\alpha_h$) is given by
\begin{equation}
K\;=\;\alpha_h\;=\;\frac{G\,M}{\sigma^2\,R_0^3}\:.
\end{equation}
Eq.\ (\ref{dPbut}) is deduced from the ``classical'' mechanical boundary
condition (Cox \cite{cox}, Eq. (17.69')), by neglecting the Eulerian variation
of the gravitational potential at the photosphere (Cowling approximation). The
advantage of Eq. (\ref{dPbut}) is that it gives $\delta P/P$, without having to
compute numerically the adiabatic or non-adiabatic eigenfunctions throughout the
entire star. Assuming that:
\begin{equation} \label{dPC}
\frac{\delta P}{P} (\theta, \phi, t) \;=\;
-\:C\:\epsilon\:P_\ell^m (\cos \theta)\:
\cos(\sigma\,t\,+\,m\,\phi)\:,
\end{equation}
we have therefore:
\begin{equation} \label{C4K}
C \;=\; 4\:+\:K^{-1} \:-\:\ell(\ell+1)\,K\:.
\end{equation}

\subsection{Influence of the effective temperature variation} 

Watson (\cite{wat}), Garrido et al.\ (\cite{gar2}) and Heynderickx et al.\ 
(\cite{hey}) related the Lagrangian temperature variation to the Lagrangian 
pressure variation at the photosphere by introduced a free {\it
ad-hoc} parameter $R$ describing the departure from adiabatic conditions. 
Moreover, they assumed that the Lagrangian temperature variation is equal
to the local effective temperature variation at the photosphere:
\begin{equation} \label{dTeffdT}
\frac{\delta T_{\rm eff}}{T_{\rm eff}}\;=\;
\frac{\delta T}{T}\:.
\end{equation}
The coefficient $f_T$ is then given by:
\begin{eqnarray} \label{Rnona}
f_T &=& R\;\frac{\Gamma_2-1}{\Gamma_2}\;|C| \nonumber \\
&=& R\;\frac{\Gamma_2-1}{\Gamma_2}\;
\big|\,4\:+\:K^{-1} \:-\:\ell(\ell+1)\,K\,\big|\:.
\end{eqnarray}
In the adiabatic case, $R=1$. Concerning the phase-lag $\psi_T$,
Heynderickx et al.\ (\cite{hey}) take the adiabatic value of 
$180^{\circ}$ for the applications to $\beta$ Cephei stars and 
Garrido et al.\ (\cite{gar2}) let $\psi_T$
be a free parameter between $90^{\circ}$ and $135^{\circ}$
for the applications to $\delta$ Scuti stars.

Our approach does not make any of these assumptions. We rigorously
compute both the amplitude and the phase of the local effective temperature 
variation by non-adiabatic computations throughout the entire star and, 
in particular, throughout the entire non-grey atmosphere.

Cugier et al.\ (\cite{cug}) and Balona \&  Evers (\cite{balev})
performed non-adiabatic computations in order to determine $f_T$ and $\psi_T$ 
in a more rigorous way. In their method, they assume also that 
the Lagrangian temperature variation is equal
to the local effective temperature variation at the photosphere 
(Eq.~(\ref{dTeffdT})).
We do not make this assumption in our method, since we have shown in Dupret et
al. (\cite{dup}) that the Lagrangian variation of the temperature at the
photosphere is different from the variation of the local effective temperature,
because of the significant Lagrangian variation of the optical depth.

\subsection{Influence of the effective gravity variation} 

Stamford \& Watson (\cite{stam}), Watson (\cite{wat}), Garrido et al.
(\cite{gar2}) and Heynderickx et al. (\cite{hey}) related the coefficient 
$f_g$ corresponding to the effective gravity 
variation (i.e.~the gravity variation corrected for the pulsational
acceleration) to the Lagrangian pressure variation by the following equation:
\begin{equation} \label{DpC}
f_g\;=\;p^*\,C\;=\;
\left(\frac{\partial\ln g}{\partial\ln P_g}\right)_{\tau=1}\:C\:,
\end{equation}
where $p^*$ is computed from equilibrium atmosphere models such 
as the models of Kurucz (\cite{kur}). Some authors 
(Cugier et al.\ \cite{cug}, Balona \& Evers \cite{balev}) proposed to 
take $p^*\,=\,1$. 

Dupret et al. (\cite{dup}, Eq. (24)) proposed a more accurate way to 
determine $\delta g_e / g_e$. In our method, $f_g$ is given by:
\begin{equation}
\label{g2}
 f_g \:=\:
   \left|\frac{R_0\,\partial\psi' / \partial r}{g\,\xi_r} 
 + \frac{4 \pi \rho\,R_0^3}{M_r} \,   
 - \left(2 + K^{-1}\right)\right|\,.  
\end{equation}
Under the Cowling approximation and neglecting the density 
at the photosphere compared to the mean density of the star, 
Eq.~(\ref{g2}) gives:
\begin{equation} \label{fgcow}
f_g\;=\; 2 \:+\: K^{-1}\;.
\end{equation}

The correction leading to Eq.\ (\ref{fgcow}) was also proposed by Cugier \&
Daszynska (\cite{cug-da}). The difference between Eq.\ (\ref{fgcow}) and Eq.\
(\ref{DpC}) with $p^*=1$ is due to the fact that the Lagrangian variation of
surface elements of the photosphere ($2-\ell(\ell+1)K$) affects the Lagrangian
variation of the pressure described in Eq.\ (\ref{C4K}), but does not affect the
Lagrangian variation of the effective gravity. A simple comparison shows that:
\begin{description}
\item[For p-modes]: $K$ is small so that the difference between 
Eq. (\ref{C4K}) and Eq. (\ref{fgcow}) is approximately 2. 
\item[For high-order g-modes]: $K$ is large, so that the difference between 
Eq. (\ref{C4K}) and Eq. (\ref{fgcow}) becomes very important~!
\end{description}
Therefore, our improvement in the determination of the effective gravity
variation has the largest impact on the photometric mode identification of 
g-mode pulsators such as SPBs and $\gamma$ Doradus stars. 

\section{Applications}

In this section, we present the application of our mode identification 
method to two SPBs and one $\beta$ Cephei star. The theoretical stellar
models we used have been computed by the Code Li\'egeois d'\'Evolution Stellaire
(CL\'ES). 

\subsection{Slowly pulsating B stars} 

The two SPBs for which we performed non-adiabatic computations and a photometric
mode identification are HD~74560 and HD~138764. Data obtained with Geneva photometry
were taken from De Cat \& Aerts (\cite{decat}). We give in Table \ref{teffloggobs} 
the effective temperature and the gravity of these two stars, 
as derived from the most recent calibration of K\"unzli et al. (\cite{kun}).
We selected then theoretical models closest to these observations and
their global characteristics are given in Tables \ref{HD74560mod} and 
\ref{HD138764mod}. Numerous additional applications to other SPBs will be presented 
in De Cat et al.\ (in preparation).

\begin{table}[!htbp]
   \caption[]{Effective temperature and gravity of the stars HD~74560 and HD~138764
              as deduced from Geneva calibrations of K\"unzli et al. (\cite{kun}).}
      \label{teffloggobs}
   \[
\begin{tabular}{lll}
   \noalign{\medskip}
     \hline
   \noalign{\medskip}
   HD~74560 & $T_{\rm eff} = 16210\:\pm\:150$\,K & $\log g = 4.15\:\pm\:0.14 $ \\
   \noalign{\medskip}
   HD~138764 & $T_{\rm eff} = 14050\:\pm\:80$\,K & $\log g = 4.20\:\pm\:0.12 $ \\
   \noalign{\medskip}
     \hline
   \noalign{\medskip}
\end{tabular}
   \]
\end{table}

\begin{table}[!htbp]
   \caption[]{Global characteristics of the theoretical model of HD~74560}
      \label{HD74560mod}
   \[
\begin{tabular}{lll}
   \noalign{\medskip}
     \hline
   \noalign{\medskip}
$M/M_{\sun} = 4.9  $  & $T_{\rm eff} = 16205  $\,K & $\log (L/L_{\sun}) = 2.7521  $ \\
   \noalign{\medskip}
$\log g = 4.1677  $  & $R/R_{\sun} = 3.0208   $    & age (My) = 24.5 \\
   \noalign{\medskip}
X = 0.7  & Z = 0.02  & no overshooting \\
   \noalign{\medskip}
     \hline
   \noalign{\medskip}
\end{tabular}
   \]
\end{table}

\begin{table}[!htbp]
   \caption[]{Global characteristics of the theoretical model of HD~138764}
      \label{HD138764mod}
   \[
\begin{tabular}{lll}
   \noalign{\medskip}
     \hline
   \noalign{\medskip}
$M/M_{\sun} = 3.9  $  & $T_{\rm eff} =  14047  $\,K & $\log (L/L_{\sun}) = 2.3760  $ \\
   \noalign{\medskip}
$\log g =  4.1964 $  & $R/R_{\sun} = 2.6073  $    & age (My) = 38 \\
   \noalign{\medskip}
X = 0.7  & Z = 0.02  & no overshooting \\
   \noalign{\medskip}
     \hline
   \noalign{\medskip}
\end{tabular}
   \]
\end{table}

\begin{table}[!htbp]
   \caption[]{Non-adiabatic results and mode identification for the star HD~74560.
              Degree $\ell$, radial order, amplitude of local effective temperature 
              variation $f_T$ and phase-lag $\psi_T$ for the modes with theoretical 
              frequency closest to the observed dominant frequency
              $f=0.64472$\,c\,d$^{-1}$.
              The identified mode is given in bold.}
      \label{HD74560res}
   \[
\begin{tabular}{llll}
   \noalign{\medskip}
     \hline
   \noalign{\medskip}
$\ell$ \hspace{0.6cm}  & $g_n$ \hspace{0.6cm}  & $f_T$ \hspace{0.6cm}  &
$\psi_T\,(\degr)$  \hspace{0.6cm} \\
   \noalign{\medskip}
{\bf 1} & $\mathbf{g_{17}}$ & {\bf 10.08} & {\bf $\!\!\!\!-$23.0} \\
2 & g$_{29}$ & 22.38 & $\!\!\!\!-8.2$ \\
3 & g$_{41}$ & 36.01 & 1.3 \\
   \noalign{\medskip}
     \hline
   \noalign{\medskip}
\end{tabular}
   \]
\end{table}

\begin{table}[!htbp]
   \caption[]{Non-adiabatic results and mode identification for the star HD~138764.
              Degree $\ell$, radial order,  $f_T$ and  $\psi_T$ 
              for the modes with theoretical 
              frequency closest to the observed dominant frequency
              $f=0.7944$\,c\,d$^{-1}$.
              The identified mode is given in bold.}
      \label{HD138764res}
   \[
\begin{tabular}{llll}
   \noalign{\medskip}
     \hline
   \noalign{\medskip}
$\ell$ \hspace{0.6cm} & $g_n$\hspace{0.6cm} & 
$f_T$ \hspace{0.6cm} & $\psi_T\,(\degr)$ \hspace{0.6cm}  \\
   \noalign{\medskip}
{\bf 1} & $\mathbf{g_{16}}$ & {\bf 5.59} & {\bf $\!\!\!\!-$23.2} \\
2 & g$_{28}$ & 17.23 & $\!\!\!\!-9.3$ \\
3 & g$_{40}$ & 31.57 & 1.4 \\
   \noalign{\medskip}
     \hline
   \noalign{\medskip}
\end{tabular}
   \]
\end{table}

\begin{figure}
  \resizebox{\hsize}{!}{\includegraphics{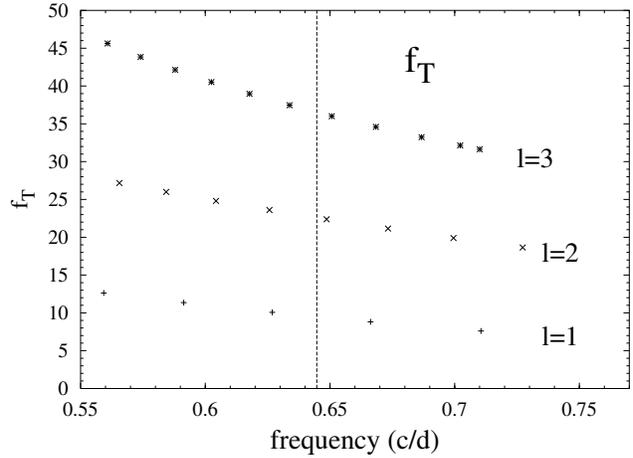}}
  \caption{$f_T$ (local effective temperature variation for a normalized
           radial displacement at the photosphere) as function of the
           pulsation frequency (in c\,d$^{-1}$), for different modes of the SPB
           star HD 74560. The ``$+$'' correspond to modes of degree $\ell=1$,
           the ``$\times$'' correspond to modes of degree $\ell=2$ and the 
           asterisks correspond to modes of degree $\ell=3$. The vertical line
           corresponds to the observed frequency of the dominant mode.}
  \label{HD74560dT}
\end{figure} 

\begin{figure}
  \resizebox{\hsize}{!}{\includegraphics{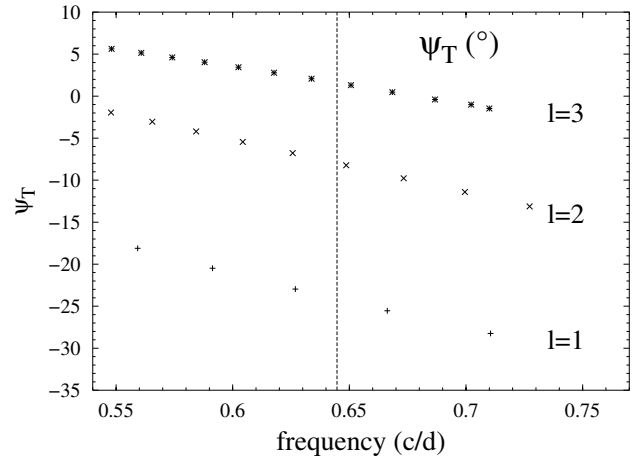}}
  \caption{$\psi_T$ (phase difference between the local effective 
           temperature variation and the radial displacement at the 
           photosphere in degrees) as function of the
           pulsation frequency in c\,d$^{-1}$, for different modes of the SPB
           star HD 74560. The ``$+$'' correspond to modes of degree $\ell=1$,
           the ``$\times$'' correspond to modes of degree $\ell=2$ and the 
           asterisks correspond to modes of degree $\ell=3$. The vertical line
           corresponds to the observed frequency of the dominant mode.}
  \label{HD74560psi}
\end{figure} 

\begin{figure}
  \resizebox{\hsize}{!}{\includegraphics{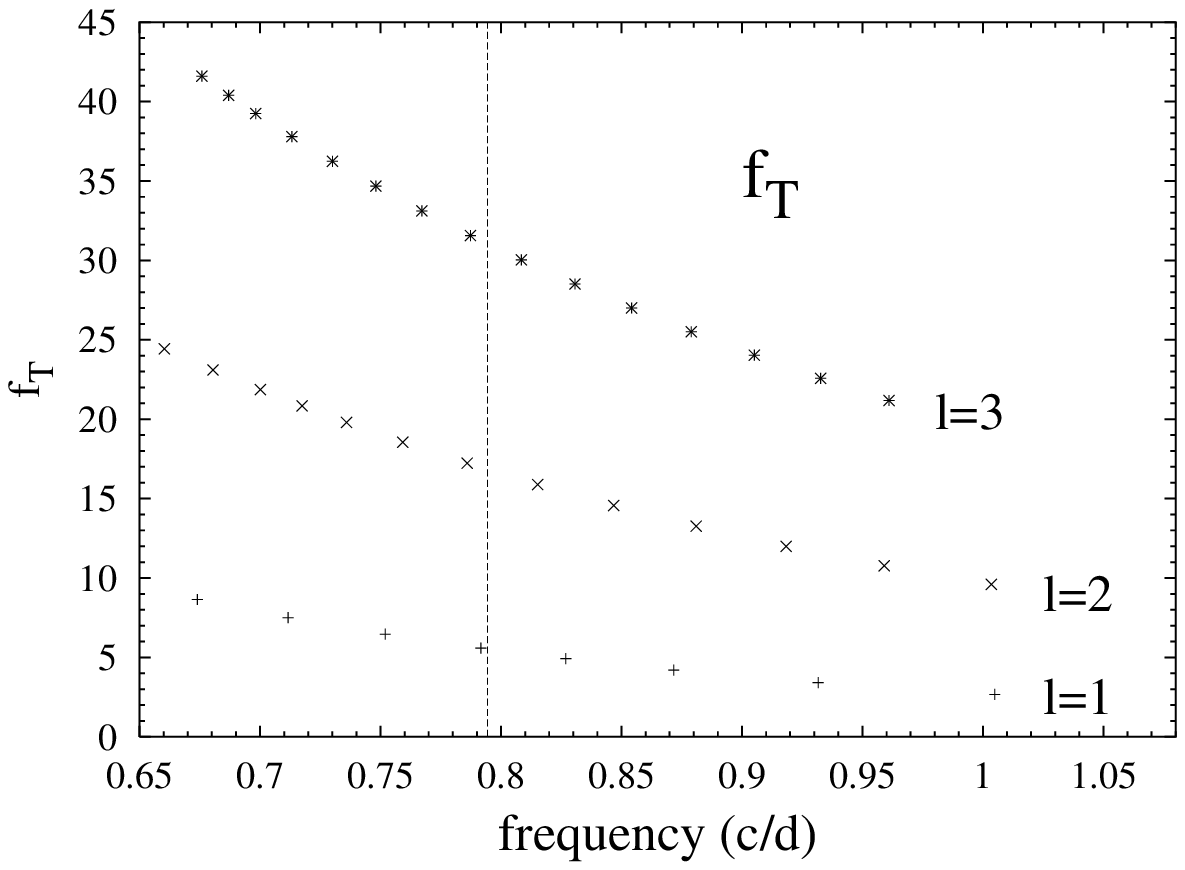}}
  \caption{$f_T$  as function of the
           pulsation frequency in c\,d$^{-1}$, for different modes of the SPB
           star HD~138764. The ``$+$'' correspond to modes of degree $\ell=1$,
           the ``$\times$'' correspond to modes of degree $\ell=2$ and the 
           asterisks correspond to modes of degree $\ell=3$. The vertical line
           corresponds to the observed frequency of the dominant mode.}
  \label{HD138764dT}
\end{figure} 

\begin{figure}
  \resizebox{\hsize}{!}{\includegraphics{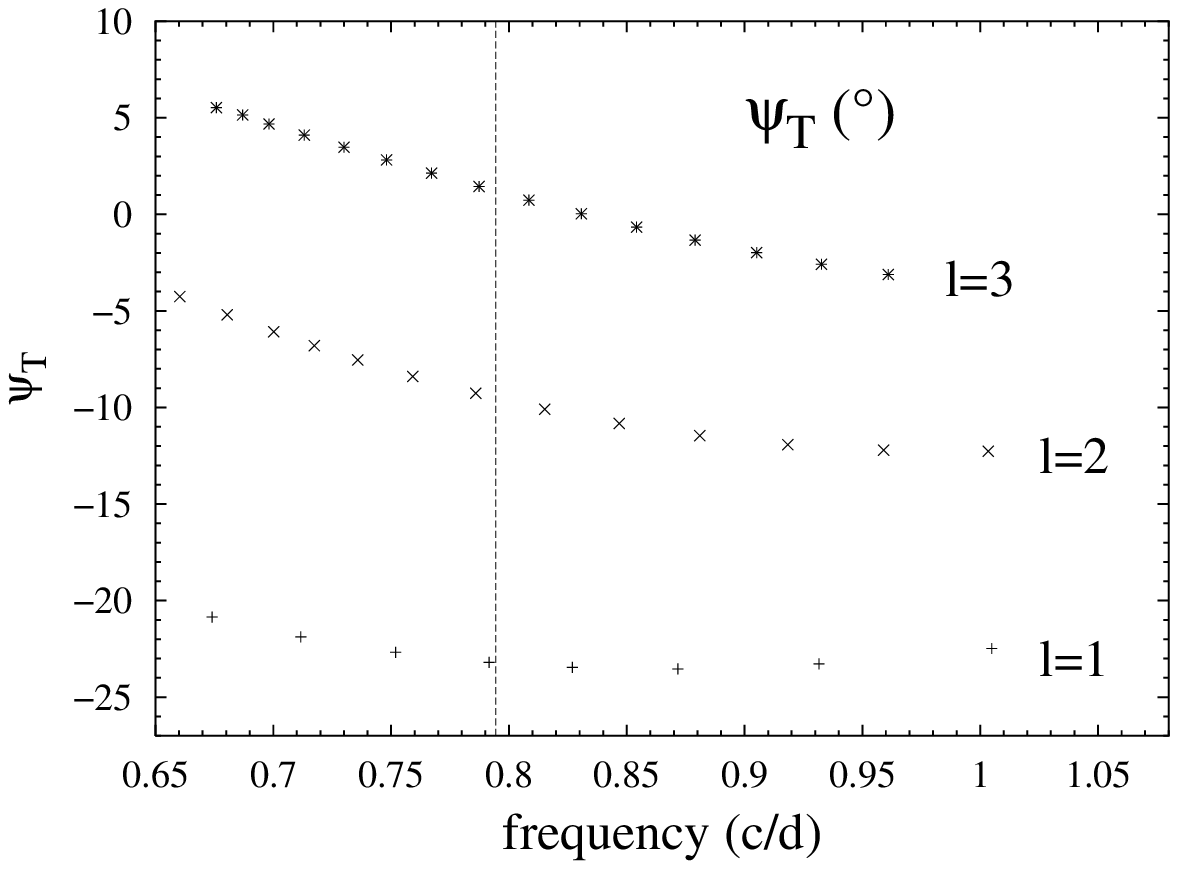}}
  \caption{$\psi_T$ (in degrees) as function of the
           pulsation frequency in c\,d$^{-1}$, for different modes of the SPB
           star HD~138764. The ``$+$'' correspond to modes of degree $\ell=1$,
           the ``$\times$'' correspond to modes of degree $\ell=2$ and the 
           asterisks correspond to modes of degree $\ell=3$. The vertical line
           corresponds to the observed frequency of the dominant mode.}
  \label{HD138764psi}
\end{figure} 

\begin{figure}
  \resizebox{\hsize}{!}{\includegraphics[angle=270]{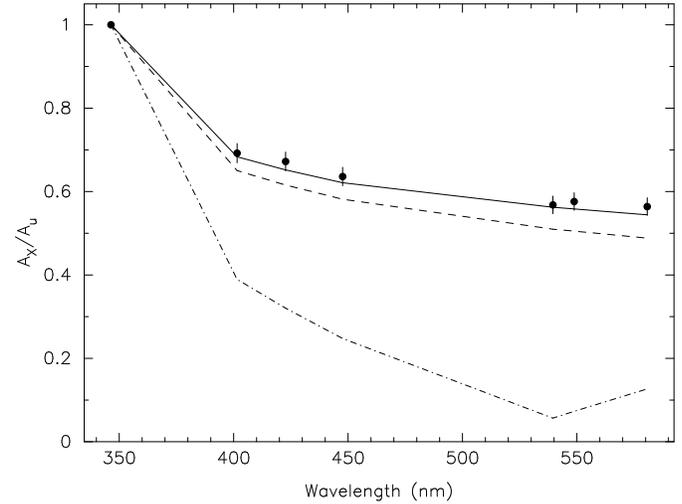}}
  \caption{Amplitude ratios obtained with Geneva photometry for the dominant mode of
  the SPB star HD~74560. The bullets with error bars correspond to the
  observations. The lines correspond to the theoretical predictions for
  different degrees $\ell$: solid line for $\ell=1$, dashed line for $\ell=2$
  and dot-dashed line for $\ell=3$.}  \label{HD74560ratio}
\end{figure} 

\begin{figure}
  \resizebox{\hsize}{!}{\includegraphics[angle=270]{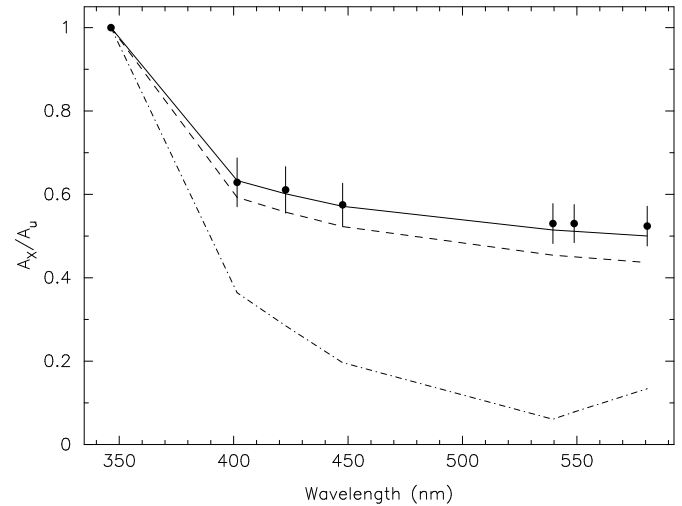}}
  \caption{Amplitude ratios obtained with Geneva photometry for the dominant mode
  of the SPB star HD~138764, same caption as in Fig.~\ref{HD74560ratio}.}  
  \label{HD138764ratio}
\end{figure} 

In Figs.~\ref{HD74560dT} and \ref{HD138764dT}, we give the values of $f_T$
(local effective temperature variation for a normalized radial displacement at
the photosphere) and in Figs.~\ref{HD74560psi} and \ref{HD138764psi} those of
$\psi_T$ (phase difference between the local effective temperature variation and
the radial displacement at the photosphere, in degrees) for different modes of the
two SPBs, as a function of the
pulsation frequency in c\,d$^{-1}$, as computed by our non-adiabatic code.  In
these figures, the vertical line corresponds to the observed frequency of the
dominant mode. 
We see that, for a given frequency, the amplitude and phase-lag
depend strongly on the degree $\ell$. The physical explanation is the
following. The term corresponding to the transversal compression in the equation
of mass conservation is proportional to $\ell(\ell+1)$. For high-order g-modes,
this term dominates, which implies a strong dependence on $\ell$ of the
eigenfunctions. The phase-lags ($\psi_T$) are relatively close to zero, which is
in agreement with the observations.

In Tables~\ref{HD74560res} and \ref{HD138764res}, we give for the two stars, the
degree $\ell$, the radial order, $f_T$ and $\psi_T$ for the modes with
theoretical frequency closest to the observed frequency. The identified mode is
given in bold.

In Figs.~\ref{HD74560ratio} and \ref{HD138764ratio}, we give the amplitude
ratios obtained from Geneva photometry for the dominant modes of the two stars.  
The bullets with error bars correspond to the observations. The full lines 
correspond to the
theoretical predictions for different degrees $\ell$: solid line for $\ell=1$,
dashed line for $\ell=2$ and dot-dashed line for $\ell=3$. For both stars, a
solution inside the error bars is found, and the identified degree is
$\ell=1$. These photometric mode identifications are in very good agreement with
the spectroscopic mode identifications performed by De Cat et al.\ (in
preparation) using the moment method.

\subsection{The $\beta$ Cephei star EN (16) Lac} 

We present now the application of our method to the $\beta$~Cephei star
EN~Lac. This star has been studied by many authors. We refer to Chapellier et al.
(\cite{chap}) and Lehmann et al.\ (\cite{leh}) for a summary of the observational
studies and to Dziembowski \& Jerzykiewicz (\cite{dziemjer}) for the first
seismic study of this $\beta$~Cephei star.  In this section, we illustrate the
process we term non-adiabatic asteroseismology, by deriving constraints on the
metallicity of the star.  The three observed frequencies used in our study were
taken from Lehmann et al.\ (\cite{leh}) and the photometric amplitudes obtained
with Johnson filters were derived by Jerzykiewicz (\cite{jerzy}).

\begin{table}[!htbp]
   \caption[]{Global characteristics of the theoretical models of EN Lac}
      \label{16lacmod}
   \[
\begin{tabular}{lll}
   \noalign{\medskip}
     \hline
   \noalign{\medskip}
Model 1a\\
   \noalign{\smallskip} 
$M/M_{\sun} = 9.4  $  & $T_{\rm eff} = 22105  $\,K & $\log (L/L_{\sun}) = 3.8992  $ \\
   \noalign{\smallskip}
$\log g = 3.8429  $  & age (My) = 16.2 & Z = 0.015\\
   \noalign{\medskip}
     \hline
   \noalign{\medskip}
Model 1b \\
   \noalign{\smallskip} 
$M/M_{\sun} = 9.7  $  & $T_{\rm eff} = 22545  $\,K & $\log (L/L_{\sun}) = 3.9405  $ \\
   \noalign{\smallskip}
$\log g = 3.8494  $  & age (My) = 15.15 & Z = 0.015\\
   \noalign{\medskip}
     \hline
   \noalign{\medskip}
Model 2a\\
   \noalign{\smallskip} 
$M/M_{\sun} = 9.5  $  & $T_{\rm eff} = 21756  $\,K & $\log (L/L_{\sun}) = 3.8769  $ \\
   \noalign{\smallskip}
$\log g = 3.8421  $  & age (My) = 15.9  & Z = 0.02 \\
   \noalign{\medskip}
     \hline
   \noalign{\medskip}
Model 2b\\
   \noalign{\smallskip} 
$M/M_{\sun} = 10  $  & $T_{\rm eff} = 22491  $\,K & $\log (L/L_{\sun}) = 3.9442  $ \\
   \noalign{\smallskip}
$\log g = 3.8548  $  & age (My) = 13.95  & Z = 0.02 \\
   \noalign{\medskip}
     \hline
   \noalign{\medskip}
Model 3a\\
   \noalign{\smallskip} 
$M/M_{\sun} = 9.7  $  & $T_{\rm eff} = 21646  $\,K & $\log (L/L_{\sun}) = 3.8739  $ \\
   \noalign{\smallskip}
$\log g = 3.8454  $  & age (My) = 14.85 & Z = 0.025\\
   \noalign{\medskip}
     \hline
   \noalign{\medskip}
Model 3b\\
   \noalign{\smallskip} 
$M/M_{\sun} = 10.3  $  & $T_{\rm eff} = 22481  $\,K & $\log (L/L_{\sun}) = 3.9532  $ \\
   \noalign{\smallskip}
$\log g = 3.8579  $  & age (My) = 12.9 & Z = 0.025\\
   \noalign{\medskip}
     \hline
   \noalign{\medskip}
\end{tabular}
   \]
\end{table}

Aerts et al.\ (\cite{aerts2}) showed
convincingly that the spectroscopic mode identification is fully compatible with
the photometric one for this star, and points towards a radial mode for the
first frequency ($f_1=5.9112$ c\,d$^{-1}$), an $\ell=2$ mode for the second
frequency ($f_2=5.8551$ c\,d$^{-1}$) and an $\ell=1$ mode for the third 
frequency ($f_3=5.5033$ c\,d$^{-1}$). 
We therefore adopt this result here.  We present the results of the amplitude
ratios obtained for models with different metallicities. The choice
of the models has been made with the following procedure. 
We computed models with 3 different metallicities~: Z = 0.015, Z = 0.02 and
Z = 0.025. Because of the uncertainties in the calibration of the effective 
temperature of EN~Lac (see Jerzykiewicz \& Sterken \cite{jerzysterk}, 
Shobbrook \cite{shob} and  Dziembowski \& Jerzykiewicz \cite{dziemjer}), 
we computed for each metallicity two evolutionary tracks with two different 
masses. 
For each evolutionary track, we subsequently selected the model giving
the best agreement between the theoretical and observed frequencies, relying on
the unambiguous mode identification. 
In all these models, X = 0.7 and there is no overshooting.
Their global characteristics are given in Table~\ref{16lacmod}.

\begin{figure}
  \resizebox{\hsize}{!}{\includegraphics{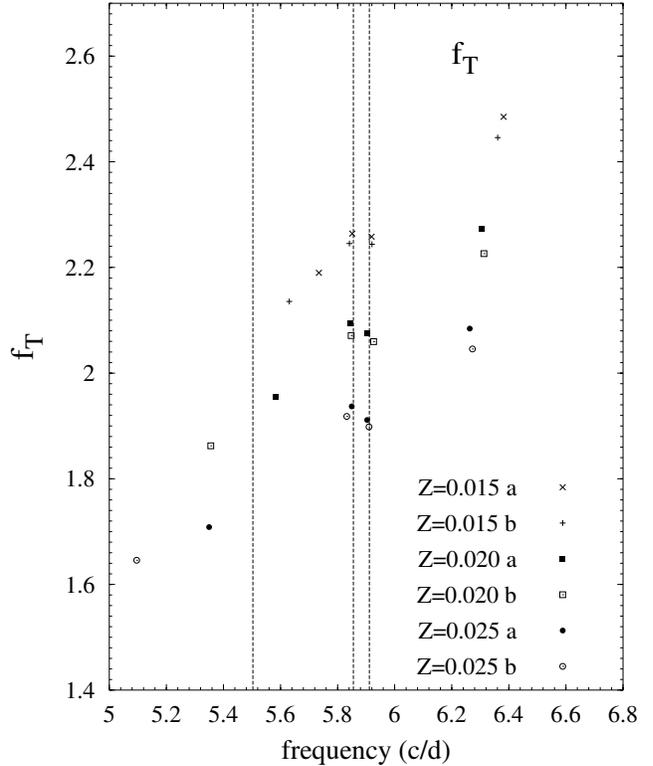}} \caption{$f_T$ (local
  effective temperature variation for a normalized radial displacement at the
  photosphere) as function of the pulsation frequency in c\,d$^{-1}$, for
  different modes ($0\,\le\,\ell\,\le\,2$) and for the six different models 
  of the star EN Lac given in Table~\ref{16lacmod}. 
  The ``$\times$''
  correspond to modes of the model 1a, the ``$+$''  correspond to the
  model 1b, the full and empty squares correspond to the 
  models 2a and 2b respectively, the full and empty circles correspond
  to the models 3a and 3b respectively. The three
  vertical lines correspond to the three observed frequencies: $\nu_1=5.9112$
  c\,d$^{-1}$, $\nu_2=5.8551$ c\,d$^{-1}$ and $\nu_3=5.5033$ c\,d$^{-1}$.}
  \label{ft16Lac}
\end{figure}

\begin{figure}
  \resizebox{\hsize}{!}{\includegraphics{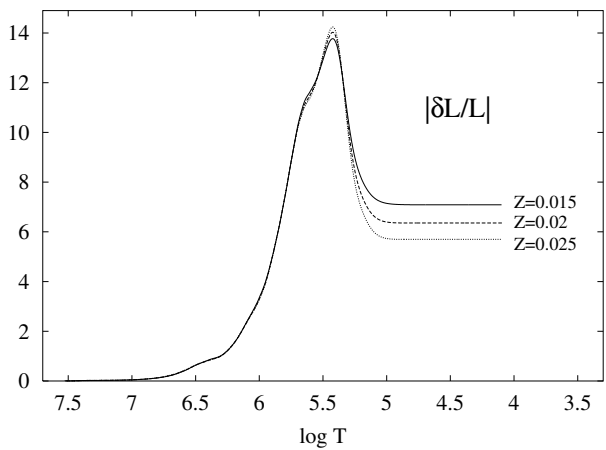}}
  \caption{Amplitude of luminosity variation $|\delta L/L|$ as function of 
           the logarithm of 
           temperature for the fundamental radial mode, for models of EN Lac
           with different metallicities (models 1a, 2a and 3a of 
           Table~\ref{16lacmod}). The solid line corresponds to the model 
           with $Z = 0.015$, the dashed line to the model with $Z = 0.02$ 
           and the dotted line to the model with $Z = 0.025$.}
  \label{dL16Lac}
\end{figure}

In Fig.~\ref{ft16Lac}, we present the values of $f_T$
as a function of the pulsation frequency in c\,d$^{-1}$, for different modes and
for the six models of EN Lac given in Table~\ref{16lacmod}. The three vertical
lines correspond to the three observed frequencies. 
We see that, the higher
the metallicity, the lower the amplitude of the local effective temperature
variation for a normalized radial displacement ($f_T$). The physical origin of this
phenomenon is explained in Fig.~\ref{dL16Lac}, where we show the amplitudes of
the luminosity variation $|\delta L/L|$ as a function of the logarithm of
temperature, from the center to the surface of the star, for the radial
fundamental mode and for the models 1a, 2a and 3a of Table~\ref{16lacmod}. The
higher the metallicity, the more efficient the $\kappa$ mechanism, which implies
a more important decrease of the luminosity variation in the driving region.
Therefore, the amplitude of the luminosity variation and of the local effective
temperature variation at the photoshere are smaller for a normalized
displacement.  The phase differences between the local effective temperature
variation and the radial displacement at the photosphere we obtained for the
different models are very close to $180^\circ$ 
(in agreement with the observations) and we do not give them here.
By comparing in Fig.~\ref{ft16Lac} the results obtained for the cold models
(models 1a, 2a and 3a) and the hot models (models 1b, 2b and 3b),
we see that changing the values of $T_{\rm eff}$ 
within its observational error bars (keeping the 
metallicity constant) has only a very small effect on the non-adiabatic results.

\begin{figure}
  \resizebox{\hsize}{!}{\includegraphics{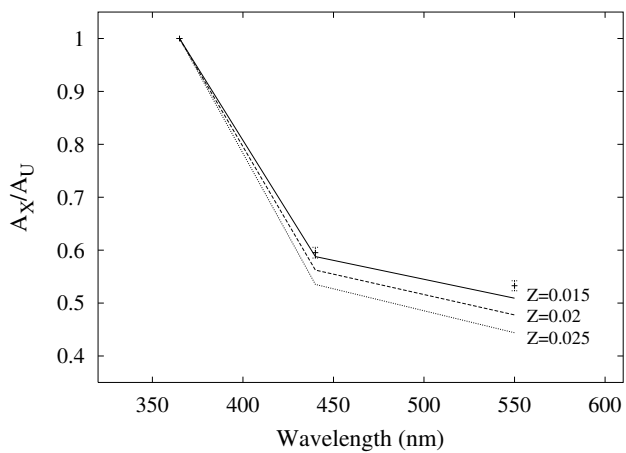}} \caption{Observed
  and theoretical amplitude ratios (Johnson photometry) for the radial
  fundamental mode obtained for three models of EN Lac 
  with different metallicities (Table~\protect\ref{16lacmod}, 
  models 1a, 2a and 3a). The solid line
  corresponds to the model with $Z = 0.015$, the dashed line to the model with
  $Z = 0.02$ and the dotted line to the model with $Z = 0.025$.
  The error bars correspond to the observations.}
  \label{16Lacdivratio}
\end{figure} 

We present in Fig.~\ref{16Lacdivratio} the theoretical amplitude ratios
(Johnson photometry) obtained for three models of EN Lac with different
metallicities (Table~\ref{16lacmod}, models 1a, 2a and 3a),  and for the 
fundamental radial mode. We see that
the model with $Z=0.015$ gives the best agreement between the theoretical and
observed amplitude ratios.  We have checked explicitly that all the modes in the
observed range of frequencies remain unstable for this low metallicity; lower
values are not compatible with mode excitation.

The confrontation between the theoretical and observed amplitude ratios
can thus be used as a constraint on the metallicity of stars driven by the metal
opacity bump ($\beta$ Cephei and Slowly Pulsating B stars), once we know
the identification of the mode.
We have seen in Fig.~\ref{ft16Lac} that the non-adiabatic predictions and
thus the theoretical amplitude ratios are little affected by the 
uncertainties on $T_{\rm eff}$ for a given metallicity, 
so that the constraints we derived on the metallicity are reliable.
This way of determining the metallicity may even turn out to be more
precise than the classical method based on the analysis of the spectrum. We plan
to validate our method to derive the metallicity by this feed-back process by
confronting our predictions to those of $\beta$ Cephei stars for which the
metallicity is known with high accuracy.

We note that, for rapidly rotating $\beta$ Cephei stars, the interaction between
pulsation and rotation can affect significantly the photometric amplitudes and
phase-lags as shown recently by Daszynska et al.\ (\cite{daszy}).  This
interaction was not yet taken into account in our current non-adiabatic
treatment.
 
\section{Conclusions} 

We have presented an improvement of the often-used photometric mode
identification method. Our version of this method is based on precise
non-adiabatic computations in which special attention is paid to the treatment
of the pulsation in the stellar atmosphere (Dupret et al. \cite{dup}).  We have
applied our new version of the method to identify the main mode of two SPBs
(HD~74560 and HD~138764) and one $\beta$ Cephei star (EN Lac, see also Aerts et
al. \cite{aerts2}).  In both cases, our photometric mode identifications were in very
good agreement with the spectroscopic mode identifications, which are far less
sensitive to temperature variations (De Ridder et al. \cite{deri}).  We have
shown also that the confrontation between the non-adiabatic theoretical
predictions and the observations can give interesting constraints on the models.
We have used the term {\it non-adiabatic asteroseismology} for this feed-back
process.  More precisely, for $\beta$ Cephei stars and SPBs, the non-adiabatic
predictions are very sensitive to the metallicity, so that this parameter can be
constrained for these stars once definite mode identification is achieved.  For
$\delta$~Scuti and $\gamma$~Doradus stars, the non-adiabatic predictions are
very sensitive to the characteristics of the thin superficial convection layer
(Balona \& Evers \cite{balev}, Moya et al. \cite{moya}, Dupret et al. \cite{dup2}). 
Therefore, it is to be
expected that a feed-back process similar to the one we presented will lead to a
significant improvement of our understanding of this convection layer.

\begin{acknowledgements}
  Part of this work was supported by the F.R.I.A. 
  (Fonds pour la formation \`a la Recherche dans
  l'Indus--trie et dans l'Agriculture). The authors are members
  of the Belgian Asteroseismology
  Group (BAG, {\tt http://www.ster.kuleuven.ac.be/$\sim$conny/bag.html}).
\end{acknowledgements}


\end{document}